\documentclass[11pt]{article}
\usepackage{moriond,epsfig}

\newcommand{\psla}{\mbox{$\not{\! p}$}}

\def\be{\begin{equation}}
\def\ee{\end{equation}}
\def\bea{\begin{eqnarray}}
\def\eea{\end{eqnarray}}

\begin{document}

\begin{flushright}
LAPTH-Conf-1255/08\\
June 2008
\end{flushright}

\vspace*{4cm}
\title{NLO QCD corrections to the production of a weak boson pair associated by a hard jet}

\author{ G. SANGUINETTI }

\address{LAPTH, Universit\'e de Savoie, CNRS, \\ 
9, Chemin de Bellevue BP 110, 74941 Annecy le Vieux, France.
\vskip 0.3 cm
{\rm S. KARG}\\
\vskip 0.1 cm
Institute for Theoretical Physics E, RWTH Aachen, D-52056 Aachen, Germany
}

\maketitle\abstracts{
In this talk we discuss recent progress concerning precise predictions for the LHC. We give a
status report of an application of the GOLEM method to deal with multi-leg one-loop amplitudes,
namely the next-to-leading order QCD corrections to the process $pp \rightarrow V V + jet$,
where $V$ is a weak boson $W^{\pm},Z$.}

The Large Hadron Collider (LHC) at CERN will probe our understanding of electroweak symmetry
breaking and explore physics in the TeV region. 
A detailed theoretical knowledge of various kinds of Standard Model backgrounds is indispensable for these studies.
Precise predictions for multi-partonic cross sections are only possible by including
higher order corrections such that renormalisation and factorisation scale dependencies are tamed. \\

During the Les Houches 2005 workshop, the process $pp \rightarrow VV jet$ was identified as one of the
most important missing next-to-leading order (NLO) calculations~\cite{LesHouches05}. 
Indeed, by considering at least one additional jet in the final state, 
one can improve the Higgs signal significance \cite{Mellado:2007fb}.
Then, this process represents an important background for the production of $H + jet$ and new physics.
Moreover, it is an important test before approaching more complicated many particle processes at NLO. \\

The process with a charged vector boson pair has
been evaluated recently by two independent groups~\cite{Dittmaier,Ellis}. 
However, the evaluation for $ZZ + jet$ is still missing.

\section{Preliminaries}
The process is composed of three partonic reactions:
\begin{equation}
q\bar{q} \rightarrow V\bar{V}g, \quad  qg \rightarrow V\bar{V}q, \quad  \bar{q}g \rightarrow V\bar{V}\bar{q}
\label{eqpp}
\end{equation}
with $V\bar{V}\in \{ZZ,W^-W^+\}$.
In fact, we only have to evaluate the helicity amplitudes of the first partonic process in (\ref{eqpp}):
\bea 
  q(p_1,\lambda_1) + \bar{q}(p_2,\lambda_2) \to V(p_3,\lambda_3) + \bar{V}(p_4,\lambda_4) + g(p_5,\lambda_5) 
\eea
Indeed, the other partonic processes are obtained by applying a momentum crossing on the partons.
For massless quarks, the allowed helicities are $\lambda_1=\lambda_2, \lambda_5\in \{-,+\}$, 
$\lambda_3,\lambda_4\in \{-,0,+\}$ which leads in general to 36 different helicity amplitudes.
The amplitude can be written as:
\bea\label{amp}
\mathcal{M}^{\lambda_1\lambda_2\lambda_3\lambda_4\lambda_5} = 
\varepsilon_{3,\mu_3}^{\lambda_3}\,
\varepsilon_{4,\mu_4}^{\lambda_4}\,
\varepsilon_{5,\mu_5}^{\lambda_5}\,
\langle 2^{\lambda_2} |\Gamma^{\mu_3\mu_4\mu_5}|  1^{\lambda_1}\rangle
\eea
where we introduce the spinor string $\Gamma^{\mu_3\mu_4\mu_5}$, 
which contains the coupling of a vector boson to a quark line given by
\bea
i\,\textrm{Vertex}^\mu_{Vq\bar{q}} &=& \, e\, \gamma^\mu ( a_{Vf\bar{f}}  - b_{Vf\bar{f}} \gamma_5 ) = \gamma^\mu ( g^+_V \Pi^+  + g^-_V \Pi^- ) 
\eea 
where $\Pi^{\pm}=(1\pm \gamma_5)/2$ and $g^\pm_V=e(a_{Vf\bar{f}}\mp b_{Vf\bar{f}})$.
We work in dimensional regularisation and treat $\gamma_5$ by applying the 't Hooft-Veltman scheme.
Therefore, the $g^-_V \, (g^+_V)$ coupling of the $Vq\bar{q}$ need a finite renormalization 
proportional to $\alpha_s \, g^+_V \, (\alpha_s\,g^-_V)$.\\

Before turning to helicity methods we notice that 
Bose symmetry~\footnote{used only for the case $V=Z$}, parity, and charge conjugation combined with parity,
relate different helicity amplitudes with each other~\footnote{See for example~\cite{DittmaierBCP}}.
Indeed, we only have to produce the following independent helicities:
\bea
\mathcal{M}_{WW}^{----\pm},\mathcal{M}_{WW}^{--00+},
\mathcal{M}_{WW}^{---0\pm},\mathcal{M}_{WW}^{---++},
\mathcal{M}_{WW}^{--0-\pm},\mathcal{M}_{WW}^{--+-+} \nonumber \\
\mathcal{M}_{ZZ}^{----\pm},\mathcal{M}_{ZZ}^{--00+},
\mathcal{M}_{ZZ}^{---0\pm},\mathcal{M}_{ZZ}^{---++} \nonumber
\eea
Then, we use these discrete transformations in order to generate the 36 helicities.\\

For the evaluation of the helicity amplitudes one preferably uses 
the spinor helicity formalism~\cite{Chinois}. Before writing down the polarisation 
vectors for the different helicities, we introduce two light-like auxiliary
vectors $k_3$, $k_4$ in order to replace the massive momenta $p_3$, $p_4$ of the $V$ bosons, 
such that $p_3+p_4 = k_3+k_4$. One finds:
\be
k_3 = \frac{1}{2\beta} [ (1+\beta)\, p_3 - (1-\beta)\, p_4 ] \quad \textrm{and} \quad
k_4 = \frac{1}{2\beta} [ (1+\beta)\, p_4 - (1-\beta)\, p_3 ] 
\ee   
where $\beta = \sqrt{1-\frac{4 M_V^2}{s_{34}}}$.
The polarization vectors for the different helicities of the massive vector bosons
can now be written now as:
\bea
{\varepsilon_{3}}^{\mu_3,+} &=& \frac{1}{\sqrt{2}}\frac{\langle 4^- |\gamma^{\mu_3}  | 3^-\rangle}{ \langle 43\rangle } \\
{\varepsilon_{3}}^{\mu_3,-} &=& \frac{1}{\sqrt{2}}\frac{\langle 3^- |\gamma^{\mu_3}  | 4^-\rangle}{ [34] } \\
{\varepsilon_{3}}^{\mu_3,0} &=& \frac{1}{2\,M_V} [ (1+\beta) {k_3}^{\mu_3} - (1-\beta) {k_4}^{\mu_3}  ]
\eea
\noindent The polarization vector ${\varepsilon_{4}}^{\mu_4,\lambda_4}$ is obtained with the relabeling $3 \leftrightarrow 4$. 
The two helicity states of the gluon are given as usual by:
\be
\varepsilon_{5\,\mu}^+ = \frac{1}{\sqrt{2}}\frac{\langle j^- |\mu  | 5^-\rangle}{ \langle j5\rangle }  \quad \textrm{and} \quad
\varepsilon_{5\,\mu}^- = \frac{1}{\sqrt{2}}\frac{\langle 5^- |\mu  | j^-\rangle}{ [5j] } 
\ee
where $j$ is a reference vector to be chosen in a convenient way. 
If $\lambda_1=\lambda_2=-1$, a convenient choice for $\mathcal{M}^{--\lambda_3\lambda_4-}$ is $j=2$
and for $\mathcal{M}^{--\lambda_3\lambda_4+}$  $j=1$.
In this way the spinor expression from the gluon can be attached to the spin chain.

By multiplying $\mathcal{M}^{--\lambda_3\lambda_4+}$ with $\langle 5^-|1|2^-\rangle/\langle 5^-|1|2^-\rangle$
and $\mathcal{M}^{--\lambda_3\lambda_4-}$ with $\langle 1^-|2|5^-\rangle/\langle 1^-|2|5^-\rangle$, 
we are now able to close the spinor string to a trace:
\bea
\mathcal{M}^{--\lambda_3\lambda_4+} &=& -
\frac{\varepsilon_{3}^{\lambda_3,\,\mu_3}\varepsilon_{4}^{\lambda_4,\,\mu_4}}{\sqrt{2} [12]\langle 1 5 \rangle^2} 
\,\,\,\textrm{tr}(\Pi^- \psla_2 \Gamma_{\mu_3\mu_4\mu_5} \psla_1 \gamma^{\mu_5} \psla_5 \psla_1) 
\label{eqGAMMAtraceplus}\\
\mathcal{M}^{--\lambda_3\lambda_4-} &=& -
\frac{\varepsilon_{3}^{\lambda_3,\,\mu_3}\varepsilon_{4}^{\lambda_4,\,\mu_4}}{\sqrt{2} \langle 1 2 \rangle [25]^2} 
\,\,\,\textrm{tr}(\Pi^- \psla_2 \, \Gamma_{\mu_3\mu_4\mu_5} \psla_1 \psla_2 \psla_5 \gamma^{\mu_5})
\label{eqGAMMAtracemoins}
\eea
In this representation it is easy to extract a global spinorial phase for each helicity amplitude.

\section{Reduction of tensor integrals}

The method used to reduce and to evaluate the one-loop tensor integrals
is the General One-Loop Evaluator for Matrix elements (GOLEM~\cite{Golem}).
This formalism is able to proceed from a Feynman diagrammatic representation of a given scattering amplitude to 
a computer code which provides a numerically stable and accurate answer for the desired cross section. \\

First we generate automatically the Feynman diagrams analytical expressions (FeynArts~\cite{FeynArts}, QGRAPH~\cite{QGraph}),
and sort the matrix element denoted $\mathcal{A}$ by helicity and colour properties.
The mapping of the diagrammatic input onto a Lorentz tensorial basis 
can be accomplished with algebraic manipulation programs (FORM~\cite{Form}).
The matrix element is then expressed in terms of some integral basis $I_B$ to be discussed below:
\begin{equation}
\mathcal{A}(|p_j \rangle,{\epsilon^{\lambda}_j},\dots) 
= \sum\limits_{BIG}  \mathcal{C}_{BIG}(s_{jk},\dots)  \times \, I_B\, \mathcal{T}_{I}(|p_j \rangle,{\epsilon^{\lambda}_j},\dots)   
\end{equation}
where the coefficients $\mathcal{C}$, depending on Mandelstam variables $s_{ij}=(p_i+p_j)^2$,
have to be simplified using algebraic programs (Maple~\cite{Maple}, Mathematica~\cite{Mathematica}).
The integral basis of our reduction algorithms only contains the scalar integrals $I^{n}_2$, $I^{n}_3$, $I^{n+2}_3$, $I^{n+2}_4$. 
For evaluating these functions, we use algebraic/numerical algorithms implemented in the flexible Fortran 90 code \texttt{GOLEM90}.

\section{Results}

With our method based on the GOLEM library, 
we have obtained the virtual order $\mathcal{O}(\alpha_s)$ corrections for all helicity amplitudes
of both processes. Using spinor helicity methods we have obtained analytical formulas for the
coefficients of all basis scalar integrals. 
In order to check the correctness of our results, we have evaluated the virtual part of both processes
twice using independent calculations and obtained full agreement.
As an illustration we show the contribution of the virtual correction to some typical distributions. 
Only the contributions which are related to finite basis integrals are plotted. 
For the full result the real emission corrections remain to be included~\cite{ZZ}.

\begin{figure}[h]
\begin{center}
\includegraphics[scale=0.55]{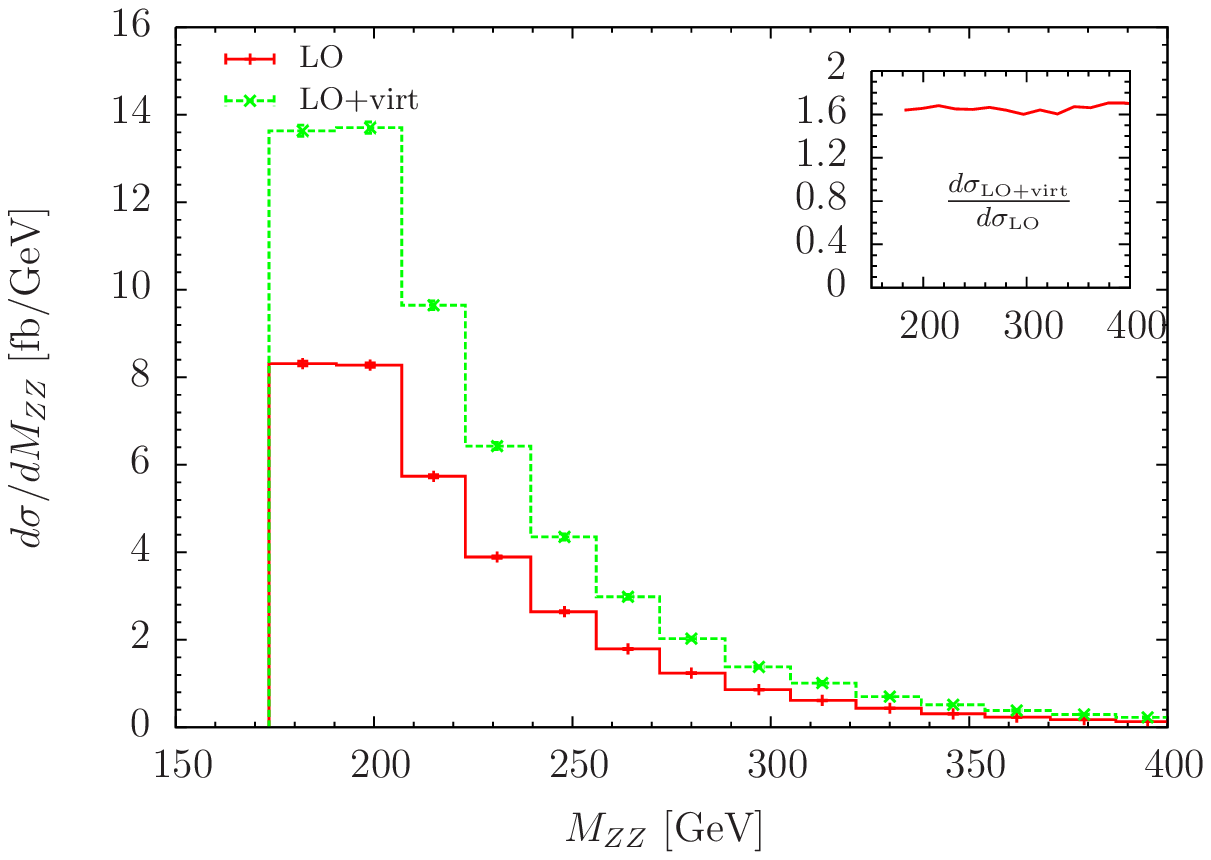}
\includegraphics[scale=0.55]{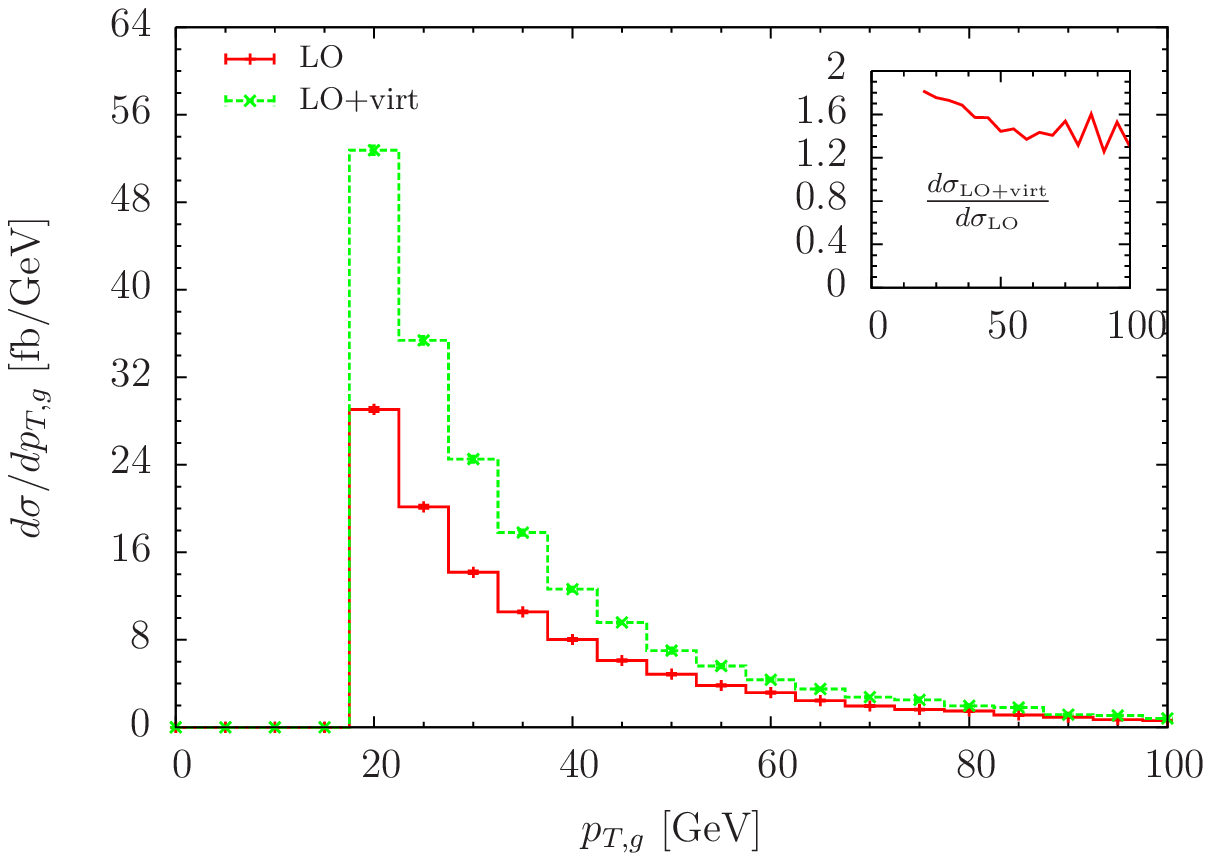}
\end{center}
\caption{\label{graphique}{The finite virtual NLO contribution to the helicity component $--+++$ of the partonic process
$q\bar{q} \rightarrow ZZg$. The invariant mass of the Z pair is shown on the left and the $p_T$ distribution on the right. We use
the cut $p_{T, jet} >$ 20 GeV and a parton and beam pipe separation cut of $\theta_{ij} > 1.50^{o}$}}
\end{figure}

\section{Tuned comparison of $pp \rightarrow W^+W^-jet$}

A tuned comparison of the NLO QCD corrections to the $W^+W^-jet$ production at the LHC 
has been done with two other groups~\cite{LesHouches07}.
We compared the integrated LO cross section and for a fixed phase-space point 
the interference term between tree-level and one-loop virtual amplitude for all channels.
The results obtained by the different groups using different calculational techniques agree within 6 to 9 digits.
The comparison of full NLO cross sections is still in progress.

\section*{Acknowledgments}
The authors would like to thank T.~Binoth, J.-P.~Guillet, and N.~Kauer. 
G.S acknowledges the conference organizing committee for financial support.


\end{document}